\documentclass[aps,pof,twocolumn,superscriptaddress]{revtex4-2}
\usepackage{graphicx}
\usepackage{epstopdf}
\usepackage{amsmath}
\usepackage{amssymb}
\usepackage{color}
\graphicspath{{figs/}}
\DeclareMathOperator{\arsinh}{arsinh}
\begin{document}
	\renewcommand{\vec}[1]{\boldsymbol{#1}}
	
	\title{Diffusioosmosis of electrolyte solutions in axisymmetric channels}
	
	\author{Elena F. Silkina}
	\affiliation{Frumkin Institute of Physical Chemistry and Electrochemistry, Russian Academy of Sciences, 31-4 Leninsky Prospect, 119071 Moscow, Russia}
	
	\author{Evgeny S. Asmolov}
	\affiliation{Frumkin Institute of Physical Chemistry and Electrochemistry, Russian Academy of Sciences, 31-4 Leninsky Prospect, 119071 Moscow, Russia}
	
	\author{Olga I. Vinogradova}
	\email[Corresponding author: ]{oivinograd@yahoo.com}
	\affiliation{Frumkin Institute of Physical Chemistry and Electrochemistry, Russian Academy of Sciences, 31-4 Leninsky Prospect, 119071 Moscow, Russia}
	
	\date{\today}
	
	\begin{abstract}
		We present a theory of fluid flow in long axisymmetric channels induced by
		salt concentration and pressure drops between their ends. The consideration
		is restricted to electrostatic diffuse layers that are thin compared to the
		local radius, but remains valid even when the concentration drop is quite large.
		We show that the magnitude of the diffusio-osmotic fluid flow rate
		$\mathcal{Q}_{\mathrm{DO}}$ in a cylinder is the same as in a slit with a
		thickness equal to the cylinder diameter, whereas channels with variable
		cross-sections can either retard or enhance it, depending on their geometry.
		The application of a pressure drop $\Delta p \neq 0$ results in an extra
		contribution $\mathcal{Q}_\mathrm{P}$ to the total fluid flow rate $\mathcal{Q}$
		without affecting $\mathcal{Q}_{\mathrm{DO}}$.
		We calculate the curves $\Delta p (\mathcal{Q})$ for several axisymmetric channels
		and conclude that they are nearly linear, with slopes that are sensitive to
		the channel shape. This leads to the possibility of introducing a simple, but
		rather accurate, cylinder approximation, where the radius of the imaginary
		cylinder is related to the hydrodynamic resistivity of the real channel and
		can be easily determined if its geometry is known. We also derive an equation
		relating the ionic flux with the total fluid flow rate, and demonstrate that
		both the sign and magnitude of the ionic flux can be tuned by using an
		appropriate channel shape. Our analysis provides a framework for interpreting
		experimental and numerical data, as well as guiding the design of micro- and
		nanofluidic devices.

	\end{abstract}
	
	\maketitle
	
	\section{Introduction}
	
	During recent decades, the rapid development of micro- and nanofluidics has
	motivated interest in interfacially driven flows~\cite{stone.ha:2004,
		squires.tm:2005, schoch2008, bocquet.l:2010}, such as electroosmosis under
	an applied electric field~\cite{gubbiotti.a:2022,vinogradova.oi:2023} or
	diffusioosmosis emerging in response to a solute gradient~\cite{anderson1989colloid}.
	The phenomenon of diffusioosmosis is particularly promising since it converts
	chemical (osmotic) energy into a directed solvent motion, leading to a variety
	of potential applications that do not require an external energy supply.
	These include separation technologies, biomedicine and healthcare, energy
	harvesting, detergency and cleaning, as well as oil recovery in porous
	media~\cite{marbach.s:2019,zhang2021,shim2022diffusiophoresis}.
	
	Diffusio-osmotic flow in micro- and nanochannels, as well as porous media,
	is a subject that currently attracts much theoretical research effort.
	Considerable attention has gone into investigating diffusio-osmotic flow
	with neutral solutes (also termed chemiosmotic flow). This topic is
	currently rather well understood, and it is widely accepted that the fluid
	velocity along the channel is proportional to the constant gradient of solute
	concentration in its ``bulk'' (central) part~\cite{anderson1982motion}.
	Thus, the problem is linear. A more comprehensive discussion can be found
	in several review articles~\cite{anderson1989colloid,marbach.s:2019}. However, all these theories become unrealistic
	for electrolyte solutions in charged channels. A steady-state diffusio-osmotic
	flow of electrolyte solutions is induced not only by a solute gradient but
	also by a tangential non-uniform electric field (the electro-osmotic
	contribution). Such an interpretation appears to date back to
	Deryagin~\cite{deryagin1961} in 1961. A corollary of this is that the local
	velocity of diffusioosmosis is proportional to the gradient of the logarithm
	of salt concentration, as shown by \citet{prieve1984motion}. Thus, the problem
	becomes non-linear even in the simplest case of a planar uniform slit.
	
	Despite numerous subsequent efforts, the theory of diffusioosmosis of
	electrolyte solutions in micro- and nanochannels remains in its infancy.
	Most analytical studies have treated the concentration gradient along the
	channel as constant~\cite{ma.hc:2006,keh.hj:2016,jing2018}, which is
	not the case unless the concentration drop is very small. A more realistic
	treatment was provided in a paper by \citet{ault.jt:2019}, who
	showed that the concentration distribution along a rectangular channel is
	generally non-linear and can be dramatically modified by the fluid flow.
	These and other authors also made important remarks about the variable surface
	potential of the channels~\cite{lee.s:2023}, but they made no attempt to
	connect the fluid flow rate with the flux of ions.
	
	Recently, we have embarked upon a detailed theoretical investigation of
	diffusio-osmotic flow for electrolyte solutions and related phenomena.
	In earlier papers, we focussed attention on slit-like channels in which
	the fluid is confined by two parallel long walls~\cite{asmolov.es:2025,
		silkina.ef:2026} and derived closed-form analytical expressions for the
	fluid flow rate, ionic fluxes, and concentration profiles. In addition,
	physically transparent approximate formulas for these quantities, as well
	as for local surface potentials, were obtained. Here, we concentrate on
	axisymmetric channels, which are generally considered to be more realistic
	models for real materials. Such channels are ubiquitous in nature
	and can be manufactured in a well-controlled way.

	Axisymmetric channels have long been considered as a crude model for real pores.
	Cylindrical, conical, and ``ink bottle'' shapes are favoured in the adsorption
	and capillary condensation literature~\cite{gregg1982adsorption} and the effect
	of geometry on these equilibrium phenomena is documented in countless publications.
	During the last decade, several theoretical papers have also been concerned with
	the role of the channel geometry in controlling various phenomena in micro- and
	nanofluidics. A large fraction of them deals with cylindrical~\cite{keh.h:2001}, conical~\cite{jubin.l:2018,boon.wq:2022,chanda.s:2022}, and corrugated~\cite{carusela.mf:2025}
	channels. For example, conical capillaries and nanopores were shown to be useful
	for current rectification~\cite{siwy.z:2006}, osmotic energy conversion~\cite{
		siria.a:2013,chang.c:2022}, and biosensing~\cite{choi.y:2006}. A few authors
	have discussed different geometries typically found in nature, from narrowing
	xylem conduits in plants~\cite{couvreur.v:2018} to ``hourglass'' shapes,
	such as those of aquaporins~\cite{gravelle.s:2013}. However, none of these
	papers addressed the issue of diffusioosmosis. The only exception is presumably the study of diffusio-osmotic flow n a cylinder
	by~\citet{keh.h:2001}, but those authors omitted important (electrostatic
	and hydrodynamic) non-linear effects to derive analytical solutions for the
	fluid velocity. These effects have recently been included by \citet{chanda.s:2022}
	in their study of diffusio-osmotic flow in conical nanochannels, which relies
	on numerical calculations. This makes the results difficult to use and limits
	predictive capabilities, although it certainly sheds some light on the flow
	in such cones. Given the current upsurge of interest in diffusioosmosis and
	its applications, it seems clear that a non-linear analytical theory for
	diffusioosmosis of electrolyte solutions in axisymmetric channels is required.
	It must provide a realistic description of the flow and have the merit of
	yielding useful (approximate) analytical results. In this paper, we develop
	such a theory by extending and generalizing an approach used
	before~\cite{asmolov.es:2025,silkina.ef:2026}.

	The remainder of the paper is arranged as follows: In Sec.~\ref{sec:model},
	we outline the model and governing equations. Section~\ref{sec:theory}
	describes the non-linear theory, providing derivations of the general
	equations for the flow rate, ionic flux, and concentration distributions.
	The results of numerical and analytical calculations of the liquid flow rate
	and ionic flux for a variety of channel geometries are then presented in
	Sec.~\ref{sec:results}. In Sec.~\ref{sec:discussion}, we introduce a simple
	cylinder approximation for channels with variable cross-sections in order to
	obtain an instructive analytical relationship between the liquid flow rate and
	pressure drop. This approximate formula provides a useful framework for
	interpreting the results of our numerical calculations. We conclude in Sec.~\ref{sec:conclusion}. Finally, the applicability limits of the linear approach are discussed in Appendix~\ref{a:1}.

	\section{Theoretical model and governing equations}\label{sec:model}
	
	\begin{figure}[h]
		\includegraphics[width=0.9\columnwidth]{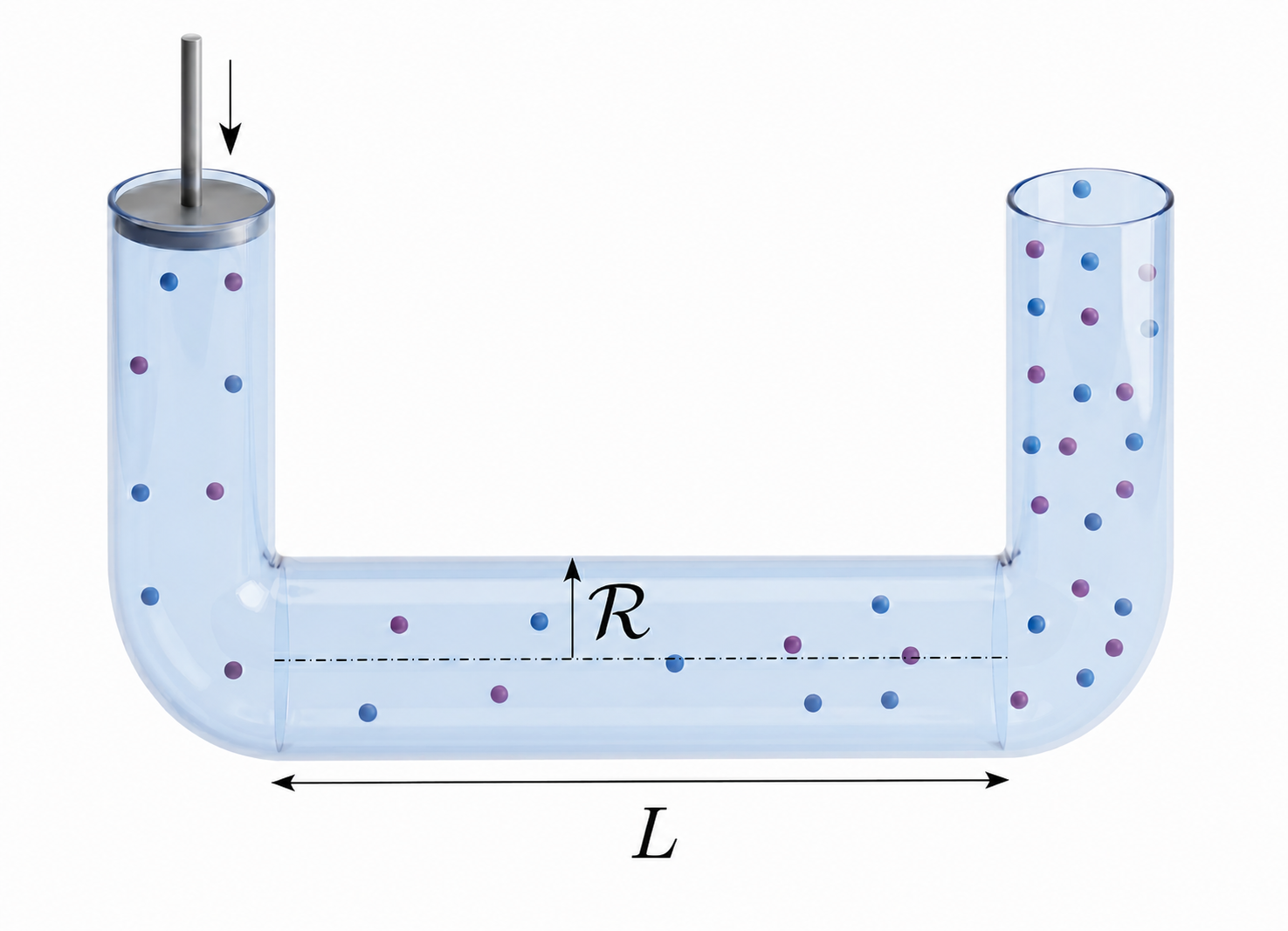}
		\caption{Sketch of an axisymmetric channel of local radius $\mathcal{R}$,
			length $L \gg \mathcal{R}$, and constant surface charge density $\sigma$
			that connects the ``fresh'' (left) and ``salty'' (right) reservoirs with
			concentrations $C_{0}$ and $C_{1}$, respectively. The hydrostatic
			pressures in these reservoirs are $P_0$ and $P_1$.}
		\label{fig:sketch}
	\end{figure}
	
\begin{figure}[h]
		\centering
		\includegraphics[width=0.9\columnwidth]{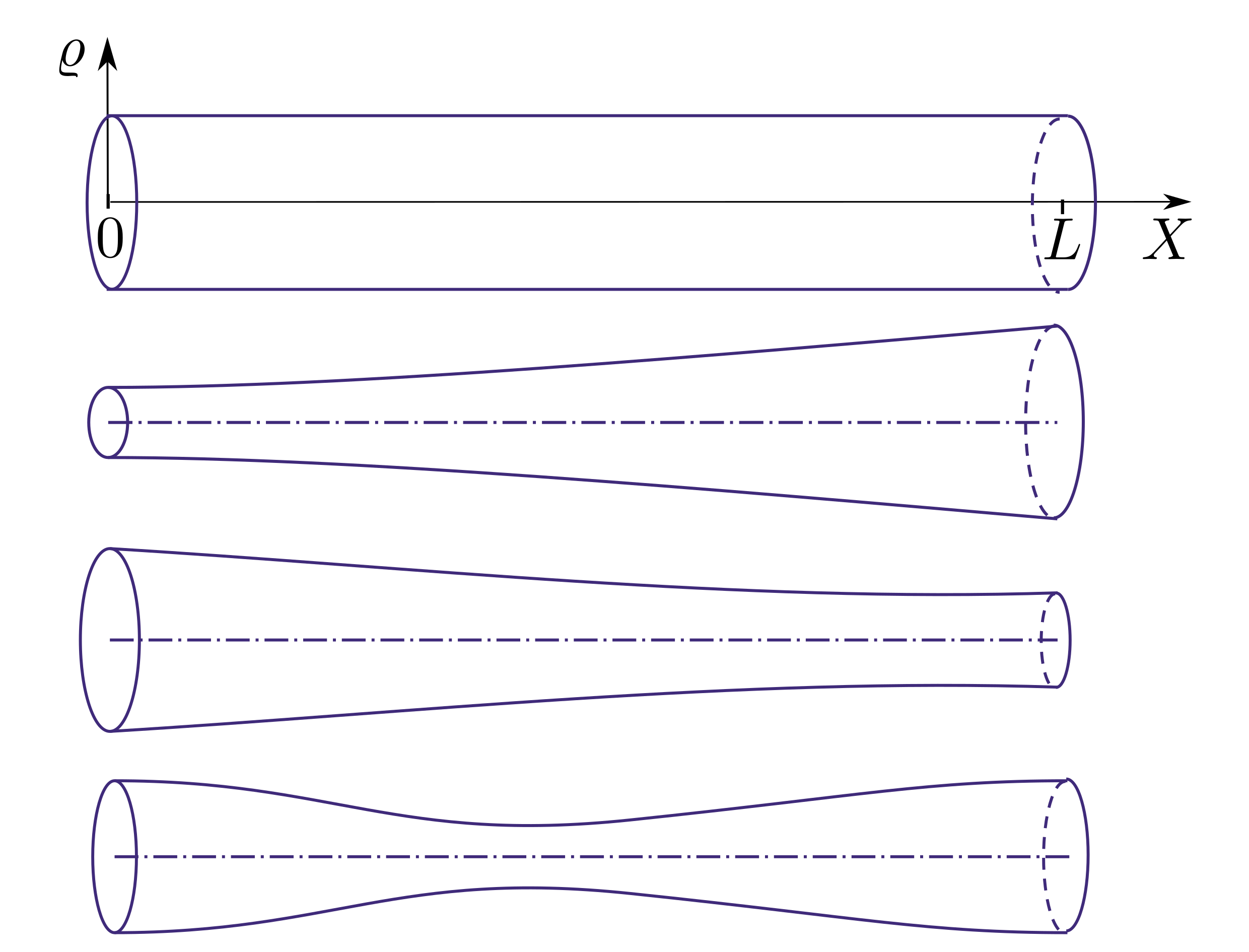}
		\caption{Schematic representation of generic axisymmetric channels connecting
			the ``fresh'' ($X=0$) and ``salty'' ($X=L$) reservoirs. From top to bottom:
			the cylinder, expanding and contracting channels (in the $X$-direction),
			and a periodically modulated pore.}
		\label{fig:channel_geometries}
	\end{figure}

	We consider a charged axisymmetric channel (cylindrical, conical, etc.) with a radius $\mathcal{R}(X)$ and length $L \gg \mathcal{R}$, as sketched in Figs.~\ref{fig:sketch} and \ref{fig:channel_geometries}. The channel is in contact with a low-salinity (left)
	and a high-salinity (right) reservoir of symmetric 1:1 salt solutions
	with bulk number densities $C_0$ and $C_1$, respectively, at standard
	temperature $T$. The cylindrical coordinate system $(X, \varrho)$ is defined
	such that the $X$-axis coincides with the channel centerline, and $\varrho$
	is the radial distance from the axis. Both electrolyte solutions have the same
	dynamic viscosity $\eta$ and permittivity $\varepsilon$. The hydrostatic
	pressures in the left and right reservoirs are $P_0$ and $P_1$.
	The pressure drop $\Delta P = P_1 - P_0$ can be of either sign or vanish.
	
	For a given radius $\mathcal{R}(X)$ and length $L$, the solution inside the
	channel will adopt a  configuration—defined by the distributions of
	the ionic number densities $C^{\pm}$, electrostatic potential $\Psi$, and
	hydrostatic pressure $P$—that minimizes the energy dissipation rate,
	yielding a steady diffusio-osmotic flow with velocity $\mathbf{U}$.
	
	The non-conducting surface is located at $\varrho = \mathcal{R}(X)$, and we
	denote its charge density as $\sigma$. Rather than using $\sigma$ explicitly,
	we describe the surface by the Gouy-Chapman length
	\begin{equation}
		\ell_{GC} = \dfrac{e}{2\pi \sigma \ell_{B}}, \label{eq:LGC}
	\end{equation}
	where $\ell_{B} = \dfrac{e^{2}}{\varepsilon k_{B}T}$ is the Bjerrum length,
	with $e$ being the elementary positive charge and $k_{B}$ the Boltzmann constant.
	The Gouy-Chapman length is inversely proportional to the surface charge density
	(and can be positive or negative depending on the sign of the charge).
	At the walls, we apply the no-slip hydrodynamic boundary condition,
	i.e., $\mathbf{U} = \mathbf{0}$ at $\varrho = \mathcal{R}(X)$.

	Before turning to the governing equations, it is instructive to ask what
	happens when the low- and high-salinity reservoirs are connected. We recall
	that the system is in a steady state, which implies that the fluid flow rate
	and ionic fluxes are constant at any cross-section. Clearly, cations and
	anions diffuse towards the ``fresh'' bath. If their diffusion coefficients,
	$D^+$ and $D^-$, are different, the cations and anions would move at different
	speeds, which is inconsistent with a steady state. Thus, in order to
	accelerate slower ions and slow down faster ones, a tangential electric field
	must spontaneously emerge. This field, in turn, induces an electro-osmotic
	flow in the channel, which supplements the standard (chemiosmotic) flow
	toward the ``fresh'' reservoir. While this physical interpretation is given
	in a similar spirit to that introduced by Deryagin~\cite{deryagin1961}
	for a single planar wall, we note that its extension to channels requires
	taking into account a non-uniform pressure gradient that emerges to maintain
	the fluid flow in each cross-section.
	
	Thus, the steady state of this complex system is governed by several
	strongly entangled phenomena. It is described by the system of
	differential equations specified below.
	
	The Nernst-Planck (or convective-diffusion) equation describes the conservation
	of ionic species at any point $(X, \varrho)$ inside the channel:
	\begin{equation}
		\mathbf{\nabla }\cdot \mathbf{J}^{\pm }=0, \label{NPH}
	\end{equation}
	where the ionic fluxes $\mathbf{J}^{\pm }$ of cations and anions are given by
	\begin{equation}
		\mathbf{J}^{\pm }=C^{\pm }\mathbf{U}+D^{\pm }\left( -\mathbf{\nabla }C^{\pm
		}\mp \dfrac{e}{k_{B}T}C^{\pm }\mathbf{\nabla }\Psi \right). \label{jd}
	\end{equation}
	In Eq.~\eqref{jd}, the first term is associated with the convective flux of ions,
	while the second and third terms represent the diffusive flux relative
	to the solvent and ionic migration in the induced electric field, respectively.

	The relation between the electrostatic potential $\Psi$ and the volume
	charge density $\rho$ is given by the Poisson equation:
	\begin{equation}
		\nabla^2 \Psi = -\frac{4\pi\rho}{\varepsilon} = -\frac{4\pi e (C^{+} - C^{-})}{\varepsilon}. \label{PEq}
	\end{equation}
	The fluid velocity and pressure satisfy the Stokes equations:
	\begin{eqnarray}
		\mathbf{\nabla \cdot U} &=& 0, \label{cont} \\
		\eta \nabla^2 \mathbf{U} - \mathbf{\nabla }P &=& \rho \mathbf{\nabla }\Psi . \label{mom}
	\end{eqnarray}
	
	The governing equations \eqref{NPH}--\eqref{mom} presented above are generic
	and apply at any point $(X,\varrho)$ inside a channel of variable radius.
	Although these equations generally require numerical methods, we demonstrate
	below that approximate analytical solutions can be derived in the
	thick-channel limit. This limit implies that a compensating charge of the
	opposite sign and equal magnitude, located in the vicinity of the charged
	walls, is confined within thin electrostatic diffuse layers near the walls,
	while an extended central region of the channel is, to the leading order,
	electro-neutral (``bulk'') in any cross-section.
	
	The local EDL thickness is of the order of the Debye screening length at a
	given cross-section and can be defined as
	\begin{equation}\label{eq:DL1}
		\lambda_{D} = \left[8\pi \ell_{B} C_m\right]^{-1/2} \propto C_m^{-1/2},
	\end{equation}
	where $C_m$ is the concentration in the electro-neutral region (equal to the
	centerline concentration).
	
	Clearly, on approaching the ``salty'' reservoir $\lambda _{D}$ reduces because the ``bulk'' concentration augments. The upper value of $\lambda_D$ is thus attained at the ``fresh'' end, where $C_m = C_0$, which yields
	\begin{equation}\label{eq:DL2}
		\lambda^{\star}_{D} = \left[8\pi \ell _{B}C_0\right] ^{-1/2}.
	\end{equation}
	Thus, to satisfy the thick channel condition, it is necessary to require $\lambda_{D}^{\star} \ll \mathcal{R}_0$, where $\mathcal{R}_0 = \mathcal{R}(0)$ is the channel radius at the ``fresh'' end.
	
	In the thick-channel limit the local electrostatic potential in the channel is given by~\cite{fair1971}
	\begin{equation}\label{eq:PSI}
		\Psi (X,\varrho) = \Psi_m (X) + \Phi (X,\varrho).
	\end{equation}
	Here the ``bulk'' term $\Psi_m$ is supplemented by a ``surface'' term $\Phi$ representing the perturbation due to diffuse layers.
	
	To construct the solution of the system of Eqs.~\eqref{NPH}--\eqref{mom}, it
	is convenient to define the dimensionless coordinates
	\begin{equation*}
		x = \frac{X}{L}, \quad s = \frac{\varrho }{\mathcal{R}_0}.
	\end{equation*}
	The dimensionless coordinate  $x$ along the channel varies from $0$ to $1$,
	and the dimensionless local channel radius is defined as
	$r(x) = \mathcal{R}(x)/\mathcal{R}_0$, so that $r_0 = r(0) = 1$.
	
	The dimensionless potentials are defined as
	\begin{equation*}
		\psi = \dfrac{e\Psi }{k_{B}T}, \quad \phi = \dfrac{e\Phi }{k_{B}T}.
	\end{equation*}
	We also introduce dimensionless variables~\cite{saville1977}
	\begin{equation*}\label{N_dim}
		\mathbf{j}^{\pm } = \mathbf{J}^{\pm }\frac{2L}{C_{0}\left( D^{+}+D^{-}\right) }, \quad
		c^{\pm }=\frac{C^{\pm }}{C_{0}}, \quad
		\mathbf{u} = \mathbf{U}\dfrac{4\pi \eta e^{2}L}{\varepsilon k_{B}^{2}T^{2}},
	\end{equation*}
	which are identical to those used in our earlier papers~\cite{asmolov.es:2025,silkina.ef:2026}.
	The dimensionless pressure is defined as
	\begin{equation*}
		p = P\dfrac{4 \pi e^{2}\mathcal{R}_0^2}{\varepsilon k_{B}^{2}T^{2}}.
	\end{equation*}
	Such a definition ensures that a given pressure drop $\Delta P$ yields
	the same $\Delta p$ as for a planar slit of half-thickness $\mathcal{R}_0$.
	
	The $x$-components of the ionic fluxes given by Eq.~\eqref{jd} can be rewritten
	in the dimensionless form as:
	\begin{equation}
		j_{x}^{\pm } = c^{\pm }\mathrm{Pe}u_x + \left( 1 \pm \beta \right)
		\left( -\partial_x c^{\pm } \mp c^{\pm } \partial_x\psi \right). \label{NP1}
	\end{equation}
	Here,
	\begin{equation}
		\mathrm{Pe} = \dfrac{k_{B} T}{2\pi \eta \ell_B \left(D^{+}+D^{-}\right)} \label{Pe}
	\end{equation}
	is the P\'eclet number, which characterizes the ratio of the rates of convection
	and diffusion. Note that $\mathrm{Pe}$ can be interpreted as an inverse effective
	diffusivity of the salt and that this (salt-specific) parameter is strictly positive. The (also salt-specific) factor $\beta$ is defined in terms of 
	the difference
	in the diffusion coefficients of cations and anions:
	\begin{equation}
		\beta = \frac{D^{+}-D^{-}}{D^{+}+D^{-}}. \label{beta}
	\end{equation}
	By definition, $\beta$ is positive if cations diffuse faster than anions,
	and negative otherwise.
	
	\section{General Analytical Framework}\label{sec:theory}
	
	\subsection{Electrostatic Potential}
	Since the channel is long, implying radial fluxes $j_{s}^{\pm} \simeq 0$,
	the ionic concentrations obey local Boltzmann distributions at any
	cross-section~\cite{fair1971,peters.pb:2016}:
	\begin{equation}
		c^{\pm } = c_m(x) \exp \left( \mp \phi \right), \label{cpm}
	\end{equation}
	where
	\begin{equation}
		\phi = \psi (x,s) - \psi_m(x) \label{fd}
	\end{equation}
	is equivalent to Eq.~\eqref{eq:PSI} rewritten in the dimensionless form.
	Here, the centerline (``bulk'') concentration $c_m$ and the potential $\psi_m$
	vary only in the $x$-direction.
	The boundary conditions for $c_m$ are
	\begin{eqnarray}\label{c0}
		c_m(0) &=& 1,\\
		c_m(1) &=& c_{1} = C_{1}/C_{0}. \label{c1}
	\end{eqnarray}
	Thus, $\Delta c = c_1 - 1$. We also set
	\begin{equation}
		\psi_m(0) = 0, \label{psi0}
	\end{equation}
	but the value of $\psi_m(1)$ that fulfills the zero-current condition
	is initially unknown and must be determined.
	
	Substituting Boltzmann distributions~\eqref{cpm} into the Poisson equation~\eqref{PEq},
	one finds that at any cross-section $\phi(s)$ satisfies the non-linear
	Poisson–Boltzmann equation:
	\begin{equation}  \label{eq:PB}
		\nabla^2 \psi = \nabla^2 \phi = \dfrac{1}{s} \dfrac{d}{d s} \left( s \dfrac{d \phi}{d s} \right) = c_m \lambda^{-2}\sinh \phi,
	\end{equation}
	where
	\[
	\lambda = \frac{\lambda_{D}^{\star}}{\mathcal{R}_0}.
	\]
	
	In the thick-channel limit, $\lambda \ll 1$, the relation between the surface
	potential and charge is given by the same equation as for a single wall
	and a wide slit~\cite{asmolov.es:2025}:
	\begin{equation}
		\phi_{s} = 2\arsinh\left( \frac{\lambda_{D}^{\star}}{c_m^{1/2}\ell_{GC}}\right). \label{eq:pot-charge_hs}
	\end{equation}

	\subsection{Global Diffusio-Osmotic Flow Rate, Ion Fluxes, and Concentrations}\label{sec:global}
	
	We focus first on the ionic fluxes $\mathcal{J}^{\pm }$ and the fluid flow
	rate $\mathcal{Q}$, which can be written in cylindrical coordinates as:
	\begin{eqnarray}	
		\mathcal{J}^{\pm } &=& 2\int_{0}^{r(x)} j_{x}^{\pm} s\, ds, \label{q} \\
		\mathcal{Q}        &=& 2\int_{0}^{r(x)} u_{x} s\, ds. \label{jq}
	\end{eqnarray}
	Here, $r(x)$ is the dimensionless local channel radius, with $r(x) \equiv 1$
	for a perfectly cylindrical channel. Due to wall impermeability,
	$\partial_{x}\mathcal{J}^{\pm } = \partial_{x}\mathcal{Q} = 0$, meaning that
	$\mathcal{J}^{\pm }$ and $\mathcal{Q}$ remain constant at any
	cross-section~\cite{asmolov.es:2025}. The main contributions to $\mathcal{J}^{\pm }$
	and $\mathcal{Q}$ in the thick-channel limit come from the central (``bulk'')
	part, where the ionic concentrations are close to $c_{m}$. Furthermore,
	the zero-current condition imposes $\mathcal{J}^{+} = \mathcal{J}^{-}$.

	The flow rate $\mathcal{Q}$ represents the sum of two components: a
	diffusio-osmotic plug-flow contribution, characterized by a constant $u_x$
	in the $s$-direction and expressed in terms of the slip velocity, and
	a pressure-driven parabolic ($u_x \propto s^2$) flow. Integrating Eq.~\eqref{jq}
	then gives
	\begin{equation}
		\mathcal{Q} = r^{2}u_{s} - r^{4}\frac{\partial_{x}p_m}{8}. \label{Q}
	\end{equation}
	Here, the local slip velocity is given by~\cite{prieve1984motion}
	\begin{equation}
		u_{s} = -\frac{\partial_{x}c_{m}}{c_{m}}\left[ \beta \phi_{s} + 4\ln \left[ \cosh \left( \frac{\phi_{s}}{4}\right) \right] \right]. \label{ui2}
	\end{equation}
	The first and second terms in Eq.~\eqref{ui2} are associated with the electro-
	and chemiosmotic slip, respectively. From Eq.~\eqref{eq:pot-charge_hs}, it follows
	that if $\lambda_{D}^{\star}/\ell_{GC}$ is prescribed, $\phi_{s}$ depends only
	on $c_{m}$, implying that $u_s$ is also a function of $c_m$ alone.
	
	Integrating the ionic fluxes given by Eq.~\eqref{NP1} over the cross-section yields:
	\begin{equation}
		\mathcal{J}^{+} \simeq c_{m}\mathrm{Pe}\mathcal{Q} + r^{2}\left[ \left( 1+\beta
		\right) \left( -\partial_{x}c_{m} - c_{m}\partial_{x}\psi_{m}\right) \right], \label{jp}
	\end{equation}
	\begin{equation}
		\mathcal{J}^{-} \simeq c_{m}\mathrm{Pe}\mathcal{Q} + r^{2}\left[ \left( 1-\beta
		\right) \left( -\partial_{x}c_{m} + c_{m}\partial_{x}\psi_{m}\right) \right]. \label{jm}
	\end{equation}
	Equations~\eqref{jp} and \eqref{jm} are applicable only if the bulk contribution to
	$\mathcal{J}^{\pm}$ dominates over the EDL contribution, which is justified
	provided that $4\ell_{Du}^{\star}/\mathcal{R}_0 \leq 1$~\cite{vinogradova.oi:2022},
	where $\ell_{Du}^{\star} = \lambda_D^{\star 2}/|\ell_{GC}|$ is the upper value
	of the Dukhin length (attained at the ``fresh'' end).

	Introducing $\mathcal{J} = \mathcal{J}^{\pm}$, one can exclude $\mathcal{Q}$
	by subtracting Eq.~\eqref{jm} from Eq.~\eqref{jp}, yielding:
	\begin{equation}
		\partial_{x}\psi_{m} = -\beta \frac{\partial_{x}c_{m}}{c_{m}}. \label{pot2}
	\end{equation}
	
	By integrating Eq.~\eqref{pot2} and imposing the boundary condition~\eqref{c0},
	we derive:
	\begin{equation}
		\psi_{m} = -\beta \ln c_{m}. \label{eq:phi0_out}
	\end{equation}
	
	It follows that $\Delta \psi = -\beta \ln c_1 \equiv -\beta \ln (1+\Delta c)$,
	i.e., by setting $\Delta c$, we simultaneously set the potential difference
	(voltage) $\Delta \psi$ between the reservoirs.
	
	Substituting Eq.~\eqref{pot2} into Eq.~\eqref{jp}, we obtain an ordinary
	differential equation:
	\begin{equation}
		\mathcal{J} = c_{m}\mathrm{Pe}\mathcal{Q} - r^2 \left( 1-\beta^{2}\right) \partial_{x}c_{m}. \label{dc}
	\end{equation}

	One additional comment should be made. The usual theory combines the global
	fluxes and driving forces in a general ``constitutive relation'':
	\[
	\begin{pmatrix}
		\mathcal{Q} \\ \mathcal{J^{+}} \\ \mathcal{J^{-}}
	\end{pmatrix}
	=
	\mathcal{M}
	\,
	\begin{pmatrix}
		\Delta p \\ \Delta \psi \\ \ln (1+\Delta c)
	\end{pmatrix},
	\]
	where $\mathcal{M}$ is a $3 \times 3$ mobility matrix (by analogy with Onsager's
	relations in bulk non-equilibrium thermodynamics~\cite{degroot_book}). This approach
	treats the matrix elements as independent of the driving forces; however, they are not.
	To see this, consider first the case of $\Delta c \ll 1$~\cite{ma.hc:2006,
		keh.hj:2016, jing2018}, which implies that variations of $c_{m}$ and $\phi_{s}$
	along the channel are small. Then, linearization gives $\partial_{x}c_{m} \simeq \Delta c$,
	$\Delta \psi \simeq -\beta \Delta c$, and $c_{m} \simeq 1$. Consequently, $u_s$
	is constant, so that $\mathcal{Q}$ and $\mathcal{J}$ can readily be calculated
	from Eqs.~\eqref{Q}--\eqref{jm}. Thus, the general ``constitutive relation''
	is applicable. However, its application to the case of $\Delta c = O(1)$
	considered here cannot be justified. The reason is that while the local linear
	relations between the fluxes and forces hold, the prefactors include the
	unknown local ``bulk'' concentration $c_m(x)$ and the surface potential $\phi_s(x)$,
	which can vary significantly. As a result, the fluxes become strongly
	non-linearly coupled to $\Delta c$, $\Delta \psi$, and $\Delta p$.
	The global flow rate $\mathcal{Q}$ and ionic flux $\mathcal{J}$ determine
	the local concentration profile $c_m(x)$, which, in turn, determines $\mathcal{Q}$.
	Such a self-consistency precludes a linear matrix representation. A quantitative
	illustration of this linear approximation breakdown is provided in Appendix~\ref{a:1}.
	
	To summarize, the self-consistent variables $\mathcal{J}$ and $\mathcal{Q}$ are
	generated because the reservoirs have different salinities and hydrostatic pressures.
	The global flow rate $\mathcal{Q}$ depends on $\Delta p$. In turn, the ionic
	flux $\mathcal{J}$ is a function of $\mathcal{Q}$. Given this entanglement,
	there are several possible choices of variables and parameters dictated,
	of course, by the physical situation. For example, in our prior work on tuning
	diffusioosmosis in slits by pressure, we fixed $\Delta p$ and
	calculated $\mathcal{Q}$~\cite{silkina.ef:2026}. However, \citet{ault.jt:2019},
	in a study of closely related problems, chose to compute $\Delta p$ by setting
	the value of $\mathcal{Q}$, rather than the other way around.
	
	The form of Eq.~\eqref{dc} suggests a solution strategy using $\mathcal{Q}$
	as a fixed input. Indeed, once $\mathcal{Q}$ and $\Delta c$ are prescribed,
	Eq.~\eqref{dc} provides a direct route to calculating $c_{m}(x)$.
	Its solution satisfying the boundary condition \eqref{c0} is
	\begin{equation}
		c_{m} = \gamma + (1 - \gamma)\exp \left( \int_{0}^{x}\frac{\alpha }{r^{2}}dx\right), \label{cm}
	\end{equation}
	with
	\begin{equation}
		\alpha = \frac{\mathrm{Pe}\mathcal{Q}}{1-\beta ^{2}}, \quad \gamma = \frac{\mathcal{J}}{\mathrm{Pe}\mathcal{Q}}. \label{gamma0}
	\end{equation}
	The value of $\alpha$ is determined immediately, since $\mathrm{Pe}$ and $\beta$
	for a specific salt are known. However, to find $\gamma$, we must first
	obtain the ionic flux $\mathcal{J}$. We return to this issue below.
	
	\subsection{Confining Geometries}\label{sec:shape}
	
	\begin{figure}[h]
		\centering
		\includegraphics[width=1.0\columnwidth]{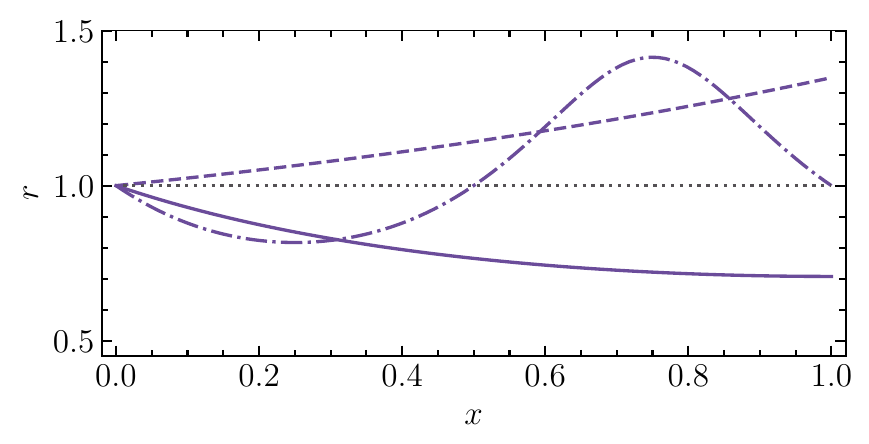}
		\caption{The profiles $r(x)$ satisfying Eq.~\eqref{rsin} that are used as
			specimen examples in this work. Dashed, dash-dotted, and solid curves
			correspond to expanding, periodically modulated, and contracting axisymmetric
			channels, respectively (see Table~\ref{table:0} for the values
			of $A$ and $\omega$). The horizontal dotted line corresponds to a
			cylindrical channel ($r=1$).}
		\label{fig:geom}
	\end{figure}
	
	The first question we address concerns the appropriate model representation of
	the confining geometries. Here, for the channel shapes $r(x)$, we propose to
	employ a family of curves:
	\begin{equation}
		r = \bigl( 1 + A\,\omega \sin(\omega x) \bigr)^{-1/2}, \label{rsin}
	\end{equation}
	where $\omega > 0$ is the wave vector of the cross-sectional disturbance
	($2 \pi / \omega \leq 1$ is its wavelength), and $A\omega$ is its amplitude.
	An obvious advantage of Eq.~\eqref{rsin} is that the integral in Eq.~\eqref{cm}
	can be taken analytically. In addition, it allows one to model various channel
	shapes, such as expanding, contracting, and periodically modulated, simply by
	choosing the appropriate values of the parameters $A$ and $\omega$. In doing so,
	we postulate that the local wall slope is small, i.e., $d\mathcal{R}/dX \ll 1$,
	allowing the problem to be tackled in the lubrication approximation.
	The criterion for its validity can be formulated as $|dr/dx| \ll L/\mathcal{R}(X)$.
	Since $L/\mathcal{R}(X)$ is large, $|dr/dx|$ can be $O(1)$ or smaller.
	Differentiating Eq.~\eqref{rsin} with respect to $x$, we find that
	$|dr/dx| \leq |A|\omega^2$. Thus, our theory applies up to $|A|\omega^2 = O(1)$.
	As a side note, the condition $\mathcal{R} \ll L$ also implies that the entrance/exit effects
	can safely be neglected, being on the order of $O(\mathcal{R}/L)$ \cite{li.z:2025}.
	
	\begin{table}[h]
		\centering
		\renewcommand{\baselinestretch}{1.2}\normalsize
		\caption{Summary of the geometric parameters used for the specimen channels.}
		\label{table:0}
		\medskip
		\begin{tabular}{|c|c|c|c|}
			\hline
			Geometry & Expanding & Sinusoidal & Contracting \\
			\hline
			$A$ & $-1.72$ & $0.08$ & $0.64$ \\
			\hline
			$\omega$ & $\pi/6$ & $2\pi$ & $\pi/2$ \\
			\hline
			$|A| \omega^2$ & $0.47$ & $3.16$ & $3.95$ \\
			\hline
		\end{tabular}
	\end{table}
	
	A limiting case of special interest is that of $A = 0$, which yields a
	cylindrical channel. For small $A \omega$, Eq.~\eqref{rsin} can be expanded
	about $r_0 = 1$ and, to the first order, reduces to the formula for a truncated
	cone. The sign of the parameter $A$ chosen for $\omega \leq \pi/2$ depends
	on whether $r$ is a monotonically increasing (minus) or decreasing (plus)
	function of $x$ in the range $0 < x < 1$. Below, we investigate both
	contracting ($A = 0.64$, $\omega = \pi/2$) and expanding ($A = -1.72$, $\omega = \pi/6$) channels. We also address a sinusoidally modulated
	($A = 0.08$, $\omega = 2\pi$) channel. The variation of their radii with $x$,
	calculated from Eq.~\eqref{rsin}, is displayed in Fig.~\ref{fig:geom}
	along with a horizontal line that corresponds to a cylinder. In addition,
	in Table~\ref{table:0}, we present the parameters in Eq.~\eqref{rsin} used
	in the present work, and demonstrate that for all these examples,
	$|A|\omega^2$ remains finite to justify the use of the lubrication approximation

	We return now to the concentration profiles. Substituting Eq.~\eqref{rsin}
	into Eq.~\eqref{cm} yields:
	\begin{equation}
		c_{m} = \gamma + (1 - \gamma) \exp \left( \alpha \left( x - A\left( \cos \left( \omega x\right)
		- 1\right) \right) \right) . \label{cm_sin}
	\end{equation}
	Using the boundary condition \eqref{c1}, we then find:
	\begin{equation}
		\gamma = \frac{c_{1} - c^{\ast }}{1 - c^{\ast }} \equiv 1 + \frac{\Delta c}{1 - c^{\ast }}, \label{gamma0_val}
	\end{equation}
	where the (positive) quantity
	\begin{equation} \label{c_star}
		c^{\ast } = \exp \left( \alpha \left( 1 - A\left( \cos \omega - 1\right) \right) \right),
	\end{equation}
	which depends on the parameters $A$ and $\omega$, increases exponentially from
	practically zero (at large negative $\mathrm{Pe}\mathcal{Q}$) to $\infty$
	as $\mathrm{Pe}\mathcal{Q} \rightarrow \infty$. Note that for the cases of
	cylindrical and sinusoidal channels, Eq.~\eqref{c_star} simplifies to:
	\begin{equation}\label{eq:cast_slit}
		c^{\ast } = \exp \left( \alpha \right),
	\end{equation}
	which is also the formula for a slit~\cite{silkina.ef:2026}. Consequently,
	$\gamma$ for cylindrical and sinusoidal channels is identical to that for a slit.
	
	Equation~\eqref{cm_sin} may be reexpressed as:
	\begin{equation}
		c_{m} = 1 + \frac{\Delta c \left[ \exp \left( \alpha \left(
			x - A\left( \cos \left( \omega x\right) - 1\right) \right) \right) - 1\right]}{c^{\ast } - 1}. \label{sm_s}
	\end{equation}
	
	Once $\gamma$ has been determined, Eq.~\eqref{gamma0} provides a direct route
	to calculating the ionic flux:
	\begin{equation}
		\mathcal{J} = \mathrm{Pe}\mathcal{Q}\left( 1 + \frac{\Delta c}{1 - c^{\ast }}\right). \label{JQ}
	\end{equation}
	While the form of this equation is identical to that derived for a slit,
	the expression for $c^{\ast }$ is, in general, different, as clarified above.
	
	We also stress that $\gamma$ becomes singular as $c^{\ast } \to 1$, which
	occurs when $\alpha \to 0$; thus, Eq.~\eqref{gamma0} implies $\mathcal{Q} \to 0$.
	By Taylor expanding Eqs.~\eqref{c_star} and \eqref{sm_s} for $|\alpha| \ll 1$,
	we find:
	\begin{equation}
		c^{\ast } \simeq 1 + \alpha\left(1 - A\left( \cos \omega - 1\right) \right), \label{c_star0}
	\end{equation}
	and
	\begin{equation}
		c_{m} \simeq 1 + \frac{\Delta c\left( x - A\left( \cos \left( \omega x\right) - 1\right) \right)}{1 - A\left( \cos \omega - 1\right) }. \label{cm0}
	\end{equation}

	From Eq.~\eqref{gamma0_val}, it follows that
	\begin{equation}
		\gamma \simeq 1 - \frac{\Delta c}{\alpha(1 - A\left( \cos \omega - 1\right)) }, \label{gamma00}
	\end{equation}
	and, using Eq.~\eqref{gamma0}, we obtain:
	\begin{equation}
		\mathcal{J} = (1 - \beta^2) \left[\alpha - \frac{\Delta c }{1 - A\left( \cos \omega - 1\right) }\right]. \label{J0}
	\end{equation}

	The pressure drop $\Delta p$ that corresponds to a given $\mathcal{Q}$ is obtained
	by integrating Eq.~\eqref{Q}, rewritten as
	\begin{equation}
		\mathcal{Q}r^{-4} = r^{-2}u_{s} - \frac{\partial_{x}p_m}{8},
	\end{equation}
	from $x = 0$ to $x = 1$, yielding:
	\begin{equation}
		\dfrac{\Delta p}{8} = \int_{0}^{1}r^{-2} u_{s} \,\mathrm{d}x - \mathcal{Q} \mathcal{W}, \label{dp}
	\end{equation}
	where $u_s$ and $r$ are given by Eqs.~\eqref{ui2} and \eqref{rsin}, respectively, and
	\begin{equation} \label{eq:definitionW}
		\mathcal{W} = \int_0^1 r^{-4} \,\mathrm{d}x.
	\end{equation}
	The first term on the right-hand side of Eq.~\eqref{dp} is associated with
	diffusioosmosis. It depends both on the shape and electrostatic properties of
	the wall, and vanishes if the wall is uncharged or $\Delta c = 0$. The second
	term is the product of the total fluid flow rate and the constant $\mathcal{W}$,
	defined solely by the channel shape. It can be easily calculated
	(analytically or numerically) if the geometry of the channel is known.
	Finally, we note that the physical meaning of $\mathcal{W}$ is apparent.
	Assume the first term in Eq.~\eqref{dp} disappears (no diffusioosmosis).
	It becomes clear that $\mathcal{W}$ characterizes the hydrodynamic resistivity
	of such a channel [usually defined as the ratio $\Delta p / \mathcal{Q}$].
	Thus, $\mathcal{W}^{-1}$ is a characteristic of hydrodynamic conductivity.

	\section{Results of calculations}\label{sec:results}
	
	In this section, we present some results for the model channels introduced
	above in Sec.~\ref{sec:shape}. As an initial application of our approach,
	we consider the diffusioosmosis of NaCl solutions, which is a commonly used
	inorganic salt characterized by $\mathrm{Pe} = 0.272$ and $\beta = -0.208$~\cite{asmolov.es:2025}.
	Since our intention here is simply to illustrate the theory, we postpone a
	comparison with other salts and its detailed discussion to a future paper.
	
	For all specimen examples below, we set $\lambda_D^{\star} = 10$~nm,
	which corresponds to a molar concentration in the ``fresh'' reservoir of
	$\mathcal{C}_0 \simeq 10^{-3}$~$\text{mol/L}$. We also fix $c_1 = 10^2$,
	which implies $\mathcal{C}_1 \simeq 10^{-1}$~$\text{mol/L}$.
	
	\subsection{Cylindrical Channels}
	
	Before describing the results of calculations for channels with variable
	cross-sections, it is instructive to consider the classical cylindrical geometry.
	
	The radius of the cylinder is constant ($r = 1$), so that Eq.~\eqref{dp} reduces to
	\begin{equation}
		\Delta p = 8\left( \int_{0}^{1} u_{s} \,\mathrm{d}x - \mathcal{Q}\right), \label{dp_cyl}
	\end{equation}
	where $u_s$ is given by Eq.~\eqref{ui2}. As explained above, the slip velocity
	depends on $c_m$ only. Therefore, one can change the variables in the integral
	and recast Eq.~\eqref{dp_cyl} as:
	\begin{eqnarray}
		\mathcal{Q} &=& -\int_{1}^{c_{1}} \frac{ \beta \phi_{s} + 4\ln \left[ \cosh \left( \phi_{s}/4\right) \right] }{c_{m}} \,\mathrm{d}c_{m} - \frac{\Delta p}{8} \label{qint}\\
		&\equiv& \mathcal{Q}_{\mathrm{DO}} + \mathcal{Q}_{\mathrm{P}}.\label{qint2}
	\end{eqnarray}
	The form of Eq.~\eqref{qint} is similar to the formula derived for a planar
	slit~\cite{silkina.ef:2026}, and the first (diffusio-osmotic) term
	$\mathcal{Q}_{\mathrm{DO}}$ is identical for both geometries. There is, however,
	an important quantitative difference between cylinders and slits hidden in the
	second (pressure-driven) term $\mathcal{Q}_{\mathrm{P}}$. In the case of a cylinder,
	the prefactor preceding $\Delta p$ is much smaller ($1/8$ versus $1/3$).
	Returning to the first term, we emphasize that the integral depends only on
	$\Delta c$ and is independent of $\mathcal{Q}$. It follows that for a cylinder,
	$\mathcal{Q}$ and $\Delta p$ are decoupled. Thus, alternatively, one may treat
	$\Delta p$ as a parameter to determine the global flow rate $\mathcal{Q}$.

	\begin{figure}[h]
		\centering
		\includegraphics[width=1.0\columnwidth]{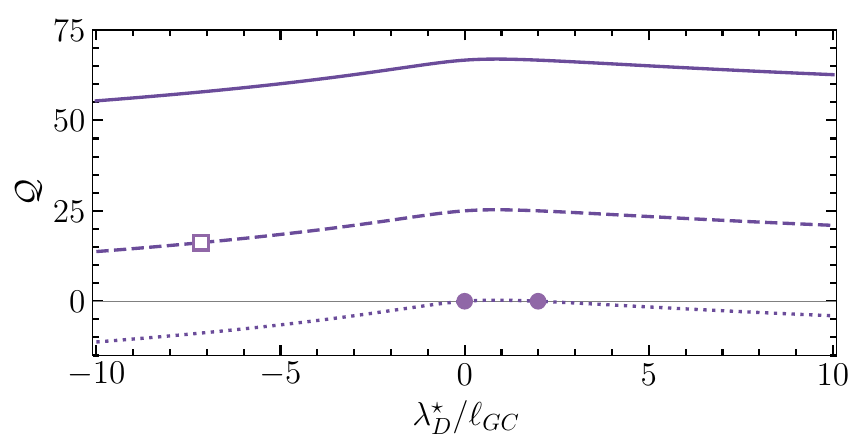}
		\caption{Fluid flow rate $\mathcal{Q}$ as a function of $\lambda_D^{\star}/\ell_{GC}$
			calculated using $\Delta p = -2 \times 10^2$ (solid and dashed curves correspond
			to the slit and the cylinder, respectively) and $\Delta p = 0$ (the dotted curve
			corresponds to both the cylinder and the slit). The filled circles and the
			open square mark the points where $\mathcal{Q} = 0$ and $\mathcal{J} = 0$, respectively.}
		\label{fig:q_cyl}
	\end{figure}

	This is illustrated in Fig.~\ref{fig:q_cyl}, where the calculations of $\mathcal{Q}$
	as a function of $\lambda_D^{\star}/\ell_{GC}\propto \sigma$ for a cylinder
	[at a given pressure drop $\Delta p$] are presented and compared with the
	results for a slit. The global flow rate $\mathcal{Q}$ is obtained by numerical
	integration of Eq.~\eqref{qint}, where the local surface potential $\phi_{s}$
	is related to the Gouy-Chapman length $\ell_{GC}$ by Eq.~\eqref{eq:pot-charge_hs}.
	When $\Delta p = 0$, the fluid flow rates in the slit and the cylinder coincide.
	For negatively charged surfaces, $\beta \phi_s$ is positive, as is the
	chemiosmotic contribution to the integral in Eq.~\eqref{qint}. Thus, $\mathcal{Q}$
	is negative, i.e., the flow is toward the ``fresh'' reservoir. As
	$\lambda_D^{\star}/\ell_{GC}$ increases, the absolute value of $\mathcal{Q}$
	decreases and becomes zero at $\lambda_D^{\star}/\ell_{GC} = 0$, that is,
	in uncharged channels. As $\lambda_D^{\star}/\ell_{GC}$ increases further
	($\beta \phi_s$ is now negative), $\mathcal{Q}$ increases, exhibits a maximum,
	and then decreases slowly, becoming, at some point, negative again.
	In Fig.~\ref{fig:q_cyl}, we have marked the points of zero flow rate with filled
	circles. This is intended to indicate the range of $\lambda_D^{\star}/\ell_{GC}$
	that corresponds to a positive branch of $\mathcal{Q}$, where the flow is toward
	the ``salty'' reservoir. We emphasize that the left circle is associated with
	zero wall charge, so diffusio-osmotic flow does not occur. The right circle
	corresponds to charged surfaces and requires an alternating integrand in
	Eq.~\eqref{qint}. Both chemi- and electroosmotic flows emerge, but their
	integral contributions cancel out. Clearly, when $\beta$ vanishes (as for
	inorganic potassium salts), the right circle disappears and no positive branch
	of $\mathcal{Q}$ exists.
	When a pressure drop of $\Delta p = -2 \times 10^2$ is applied, the (dotted)
	curve for both geometries is shifted to larger $\mathcal{Q}$ by an amount
	proportional to $\Delta p$; thus, the qualitative features of diffusioosmosis
	in the slit and cylinder are identical. The quantitative difference between
	these two geometries is that this shift is much smaller for a cylinder.
	Finally, we remark that the shift can reverse the flow. This is clearly seen
	in Fig.~\ref{fig:q_cyl}, where the effect of a negative $\Delta p$, coupled to
	the diffusio-osmotic contribution, results in a flow toward the ``salty''
	reservoir for the entire range of $\lambda_D^{\star}/\ell_{GC}$ shown here.
	
	\begin{figure}[h]
		\centering
		\includegraphics[width=1.0\columnwidth]{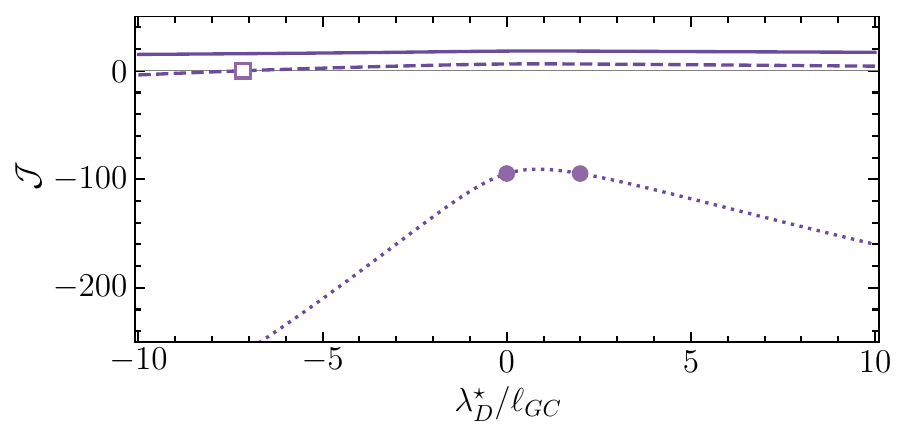}
		\caption{Ionic flux $\mathcal{J}$ as a function of $\lambda_D^{\star}/\ell_{GC}$,
			calculated for the same parameters as in Fig.~\ref{fig:q_cyl}.}
		\label{fig:j_cyl}
	\end{figure}

	By using Eq.~\eqref{JQ} with $c^{\ast}$ given by Eq.~\eqref{eq:cast_slit}, one
	can obtain the curves for $\mathcal{J}(\lambda_D^{\star}/\ell_{GC})$ displayed
	in Fig.~\ref{fig:j_cyl}. All of them exhibit a maximum precisely at the
	maximum of $\mathcal{Q}$. When $\Delta p = 0$ (the lowermost curve),
	$\mathcal{J}$ is negative for all $\lambda_D^{\star}/\ell_{GC}$, suggesting
	that the diffusio-osmotic flow tends to equalize the salt concentrations
	on the two sides. If $\mathcal{Q}$ vanishes, Eq.~\eqref{J0} is invoked to
	calculate the ionic flux. For a cylinder, this reduces to:
	\begin{equation}
		\mathcal{J} \bigl|_{\mathcal{Q} = 0} \,\simeq -(1 - \beta^2)\Delta c,
	\end{equation}
	which is identical to the formula for a slit derived before~\cite{asmolov.es:2025, silkina.ef:2026}.
	For our specimen example, this simple equation yields $\mathcal{J}\bigl|_{\mathcal{Q} = 0} \simeq -94.72$.
	The reason for the negative $\mathcal{J}$ at zero $\mathcal{Q}$ is the combined
	effect of ion diffusion toward the ``fresh'' reservoir and their migration
	caused by the induced voltage $\Delta \psi$.
	
	When a negative pressure drop is applied, $\mathcal{Q}$ and $\alpha$ become
	positive. The maximum of $\mathcal{J}$ is much less pronounced and is
	practically invisible on the scale of Fig.~\ref{fig:j_cyl}. In the case of a
	slit (the uppermost curve), $\mathcal{J}$ is positive for all
	$\lambda_D^{\star}/\ell_{GC}$ shown, whereas as $\lambda_D^{\star}/\ell_{GC}$
	increases, the ionic flux in the cylinder changes its sign from negative to positive.
	The reason for this difference is clear. From Eq.~\eqref{JQ}, it follows that
	the point of zero $\mathcal{J}$ is attained when
	\begin{equation}\label{eq:zeroJcyl}
		\mathcal{Q}\bigl|_{\mathcal{J} = 0} \, = \dfrac{ (1-\beta^2) \ln c_1}{\mathrm{Pe}}.
	\end{equation}
	With our parameters, this yields $\mathcal{Q}\bigl|_{\mathcal{J} = 0} \simeq 16.2$.
	The values of $\mathcal{Q}$ in the slit for $-10 \leq \lambda_D^{\star}/\ell_{GC} \leq 10$
	are above this value, but in the cylinder, $\mathcal{J} = 0$ is attained at
	$\lambda_D^{\star}/\ell_{GC} \simeq -7.2$. This point of zero ionic flux is
	indicated in Fig.~\ref{fig:j_cyl} by the open square and is also plotted in
	Fig.~\ref{fig:q_cyl}.

	\subsection{Channels of Varying Cross-Section}\label{sec:resultsB}
	
	For channels with variable $r$, a different story obtains. Substituting
	Eq.~\eqref{rsin} into Eq.~\eqref{eq:definitionW} and performing the integration,
	one can easily derive:
	\begin{equation}\label{eq:int_r4}
		\mathcal{W} = 1 + 2A(1 - \cos\omega) + \frac{A^2\omega^2}{2} - \frac{A^2\omega\sin(2\omega)}{4}.
	\end{equation}
	This general expression can be significantly simplified for the specific
	geometries listed in Table~\ref{table:0}. For example, for the sinusoidal channel,
	Eq.~\eqref{eq:int_r4} reduces to $\mathcal{W} = 1 + A^2\omega^2/2$.
	We emphasize that this formula is different from $\mathcal{W} = 1$ for a cylinder.
	This immediately suggests that the relation between $\Delta p$ and $\mathcal{Q}$
	is also modified, whereas the expression for $\mathcal{J}$ remains the same
	[see Sec.~\ref{sec:shape}].
	
	The first integral on the right-hand side of Eq.~\eqref{dp}, however, cannot
	be re-expressed solely in terms of $c_m$, as in Eq.~\eqref{qint}, since the
	integrand depends on $r$ as well. This implies that the integral itself is a
	function of both $\Delta c$ and $\mathcal{Q}$; therefore, it is more convenient
	to prescribe $\mathcal{Q}$ and then determine $\Delta p$ from Eq.~\eqref{dp}.
	
	Next, we calculate the global channel properties, such as the pressure drop
	required to maintain a prescribed flow rate and the corresponding ionic flux.
	The calculations are performed using a fixed value of $\lambda_D^{\star}/\ell_{GC} = -10$,
	which corresponds to $\ell_{GC} = -1$~nm. Note that this value of $\ell_{GC}$
	matches the maximum surface charge density reported in experiments~\cite{stein.d:2004,
		kitamura.a:1999,cerovic.ls:2002}.
	
	\begin{figure}[h]
		\centering
		\includegraphics[width=1.0\columnwidth]{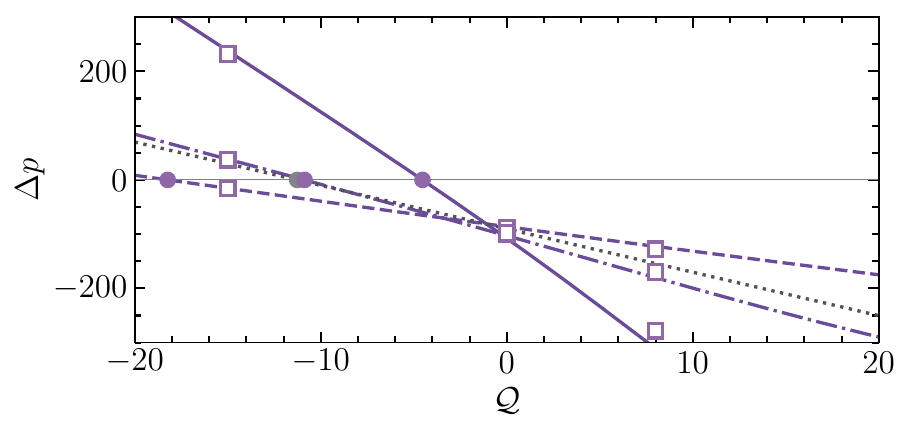}
		\caption{Pressure drop $\Delta p$ as a function of the flow rate $\mathcal{Q}$.
			The dotted straight line corresponds to the analytical calculations for a cylinder.
			Solid, dash-dotted, and dashed curves correspond to contracting,
			sinusoidally modulated, and expanding axisymmetric channels, respectively,
			as specified in Fig.~\ref{fig:geom}. The filled circles correspond to
			$\mathcal{Q}_{\mathrm{DO}}$ from Table~\ref{table:1}, while the open squares
			represent $\Delta p$ calculated from Eq.~\eqref{dp2}.}
		\label{fig:dp_vs_Q}
	\end{figure}
	
	Figure~\ref{fig:dp_vs_Q} includes theoretical curves for $\Delta p (\mathcal{Q})$
	obtained by numerical integration of Eq.~\eqref{dp} for several geometries
	defined in Sec.~\ref{sec:shape}. It can be seen that as $\mathcal{Q}$ increases,
	$\Delta p$ decreases monotonically for all channels. In essence, on the scale
	of this figure, all the curves appear to be straight lines with different slopes,
	but in reality, only the curve for the cylinder is strictly linear.
	There is some deviation from linearity for the other curves, which, however,
	can be treated as insignificant. We return to this issue in Sec.~\ref{sec:discussion}.
	\begin{figure}[h]
		\centering
		\includegraphics[width=1.0\columnwidth]{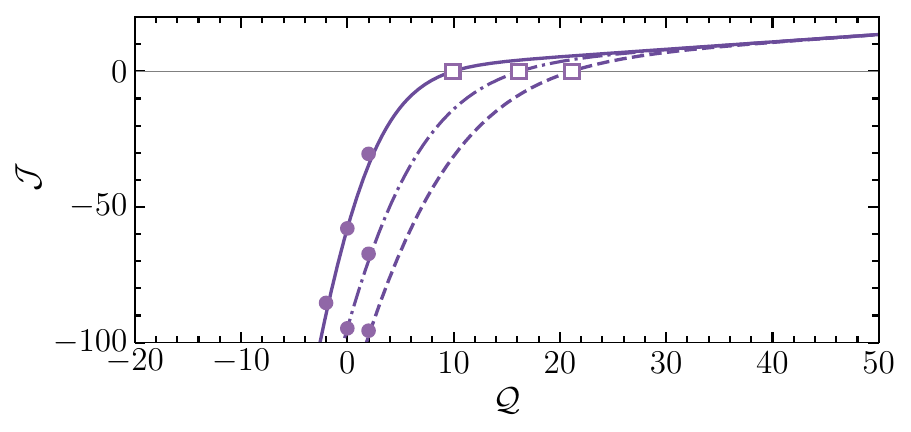}
		\caption{Ionic flux $\mathcal{J}$ as a function of the total fluid flow rate $\mathcal{Q}$.
			Solid, dash-dotted, and dashed curves correspond to contracting,
			sinusoidally modulated (identical to cylindrical), and expanding axisymmetric
			channels, respectively. The filled circles represent the calculations from
			Eq.~\eqref{J0}, while the open squares mark the points where $\mathcal{J} = 0$.}
		\label{fig:j_pe}
	\end{figure}
	
	In Fig.~\ref{fig:j_pe}, we plot $\mathcal{J}$ calculated as a function of
	$\mathcal{Q}$ from Eq.~\eqref{JQ}. Note that the ionic flux in the cylinder
	is not shown because it is identical to that of the sinusoidal channel.
	It can be seen that when $\mathrm{Pe}\mathcal{Q} \gg 1$, all curves virtually
	merge into a single straight line. In this limit, $\gamma \simeq 1$ and the
	ionic flux $\mathcal{J} \simeq \mathrm{Pe}\mathcal{Q}$ is toward the ``salty''
	reservoir. If $\mathcal{Q}$ is reduced, the curves calculated for different
	channel shapes separate. As a result, their intersections with the horizontal
	axis $\mathcal{J} = 0$ [marked with open squares] correspond to different
	(positive) values of $\mathcal{Q}$. These can be determined from Eq.~\eqref{gamma0_val}
	by setting $c^{\ast} = c_1$, which yields:
	\begin{equation}\label{eq:zeroJQ}
		\mathcal{Q}\bigl|_{\mathcal{J} = 0} \, = \dfrac{(1-\beta^2)\ln c_1}{\mathrm{Pe} \left[1-A\left( \cos \omega -1\right)\right]}.
	\end{equation}
	As expected, in the case of the sinusoidal channel, this equation reduces
	to Eq.~\eqref{eq:zeroJcyl} for a cylinder. However, the flow rate $\mathcal{Q}$
	yielding zero $\mathcal{J}$ is smaller for the contracting ($A > 0$) channel,
	and larger for the expanding ($A < 0$) channel. On reducing $\mathcal{Q}$
	further, $\mathcal{J}$ decreases rapidly (increases in magnitude), and when
	$\mathrm{Pe} |\mathcal{Q}| \leq 1$, it is well described by Eq.~\eqref{J0}.
	In the limit of large negative $\mathrm{Pe}\mathcal{Q}$, the curves again
	collapse into a single straight line (not shown). In this situation,
	$\gamma \simeq c_1$, and thus $\mathcal{J} \simeq \mathrm{Pe}\mathcal{Q}c_1$.

	\section{The Cylinder Approximation and Its Consequences}\label{sec:discussion}

	Several of the results obtained from the calculations in the preceding section
	can be understood in terms of a cylinder approximation to the channels with
	variable cross-sections. Indeed, the apparent linearity of the curves
	$\Delta p(\mathcal{Q})$ in Fig.~\ref{fig:dp_vs_Q} implies that the relation
	between $\Delta p$ and the global flow rate $\mathcal{Q}$ is approximately
	the same as for a cylinder of an effective radius $\mathcal{R}_{\mathrm{eff}} \neq \mathcal{R}_0$.
	This radius can be evaluated by equating $\int_{0}^{1}r_{\mathrm{eff}}^{-4}\,\mathrm{d}x = r_{\mathrm{eff}}^{-4}$
	to $\mathcal{W}$ given by Eq.~\eqref{eq:int_r4}, which yields:
	\begin{equation}
		r_{\mathrm{eff}} = \dfrac{\mathcal{R}_{\mathrm{eff}}}{\mathcal{R}_0} = \mathcal{W}^{-1/4}. \label{reff}
	\end{equation}
	Dividing then Eq.~\eqref{dp} by $\mathcal{W}$, one can recast it to a form
	similar to Eq.~\eqref{qint2} for a cylinder:
	\begin{equation}
		\mathcal{Q} \simeq \mathcal{Q}_{\mathrm{DO}} - \dfrac{r_{\mathrm{eff}}^{4}\Delta p}{8}, \label{dp2}
	\end{equation}
	where the diffusio-osmotic term
	\begin{equation}
		\mathcal{Q}_{\mathrm{DO}} = \mathcal{Q} \bigl|_{\Delta p = 0} = \frac{\int_{0}^{1}r^{-2} u_{s} \,\mathrm{d}x}{\mathcal{W}}, \label{qeff}
	\end{equation}
	may be determined numerically from $\Delta p(\mathcal{Q}) = 0$. This needs to be
	calculated only once, since $\mathcal{Q}_{\mathrm{DO}}$ represents a constant
	independent of $\Delta p$. We emphasize that from Eq.~\eqref{dp2}, it is almost
	self-evident that this quantity depends on the channel shape, mostly due to
	the differences in hydrodynamic resistivity of the channel, which also defines
	the slope of the curve $\Delta p(\mathcal{Q})$.
	
	\begin{table}[h]
		\centering
		\renewcommand{\baselinestretch}{1.2}\normalsize
		\caption{Values of $\mathcal{W}$, effective radius $r_{\mathrm{eff}}$, diffusio-osmotic
			contribution $\mathcal{Q}_{\mathrm{DO}}$ to the flow rate, and the pressure drop
			$\Delta p$ at zero $\mathcal{Q}$, calculated using the parameters from Table~\ref{table:0}.}
		\label{table:1}
		\medskip
		\begin{tabular}{|c|c|c|c|c|}
			\hline
			Geometry & Expanding & Cylinder & Sinusoidal & Contracting \\
			\hline
			$\mathcal{W}$ & 0.61 & 1.00 & 1.13 & 2.77 \\
			\hline
			$r_{\mathrm{eff}}$ & 1.13 & 1.00 & 0.97 & 0.77 \\
			\hline
			$\mathcal{Q}_{\mathrm{DO}}$ & $-18.23$ & $-11.26$ & $-10.87$ & $-4.53$ \\
			\hline
			$\Delta p \bigl|_{\mathcal{Q} = 0}$ & $-89.45$ & $-90.08$ & $-97.83$ & $-100.45$ \\
			\hline
		\end{tabular}
	\end{table}

	To provide concrete values, Table~\ref{table:1} presents $\mathcal{W}$,
	$r_{\mathrm{eff}}$, and $\mathcal{Q}_{\mathrm{DO}}$ calculated from
	Eqs.~\eqref{eq:int_r4}, \eqref{reff}, and \eqref{qeff}, respectively.
	The data are listed in decreasing order of the effective radius and the
	magnitude of the negative diffusio-osmotic contribution to $\mathcal{Q}$.
	
	The intersections of the horizontal line $\Delta p = 0$ with the corresponding
	numerical curves in Fig.~\ref{fig:dp_vs_Q} determine $\mathcal{Q}_{\mathrm{DO}}$.
	The data from Table~\ref{table:1} are also plotted as filled circles, thereby
	confirming the validity of the effective-cylinder model. Note that the magnitudes
	of $\mathcal{Q}_{\mathrm{DO}}$ are very close, as are the effective radii of
	the cylindrical and sinusoidal channels.
	The expanding (toward the ``salty'' reservoir) channel is different.
	Since the flow is directed toward the ``fresh'' end, this channel narrows
	in the direction of the fluid motion and accelerates it. Conversely,
	the contracting channel widens in the diffusio-osmotic flow direction,
	slowing down the flow rate.

	Note that Eq.~\eqref{dp2} can be used to find $\Delta p$ at zero $\mathcal{Q}$:
	\begin{equation}
		\Delta p \bigl|_{\mathcal{Q} = 0} \simeq \dfrac{8 \mathcal{Q}_{\mathrm{DO}}}{r_{\mathrm{eff}}^{4}}. \label{dpzq}
	\end{equation}
	It seems clear that to block the liquid flow across the channel, $\Delta p$ and
	$\mathcal{Q}_{\mathrm{DO}}$ must be of the same sign. Specializing to NaCl,
	this implies that $p_0 > p_1$, meaning that $\Delta p < 0$. The values of
	$\Delta p \bigl|_{\mathcal{Q} = 0}$ are also included in Table~\ref{table:1}.
	At first sight, it is somewhat surprising that a smaller $|\mathcal{Q}_{\mathrm{DO}}|$
	requires a larger $|\Delta p|$. However, it becomes almost self-evident from
	the values of $r_{\mathrm{eff}}$ summarized in Table~\ref{table:1} that this is
	a consequence of the smaller cross-sectional area of the effective cylinder,
	leading to enhanced hydrodynamic resistance to the pressure-driven flow.
	
	The impact of $\Delta p$ on the total flow in channels with different shapes
	deserves a more detailed discussion. If we apply a positive $\Delta p$,
	or a negative $\Delta p$ that is greater than the pressure of zero flow rate,
	the qualitative picture remains the same as that observed for a pure
	diffusio-osmotic flow. Namely, $\mathcal{Q}$ is negative, and its magnitude
	is largest for a channel that narrows in the direction of the flow.
	There exists, of course, an important quantitative difference, as is clearly
	seen in Fig.~\ref{fig:dp_vs_Q}.
	If we now set $\Delta p < 0$ such that it is less than $\Delta p \bigl|_{\mathcal{Q} = 0}$,
	the positive fluid flow rate increases according to the sequence:
	the contracting channel $<$ sinusoidal $<$ cylindrical $<$ expanding.
	Thus, $\mathcal{Q}$ reduces in a channel that widens in the stream direction,
	and augments in a channel that narrows. This conclusion is exactly the opposite
	of that made for a positive $\Delta p$. Presumably, this difference is due
	to the dominant contribution of $\mathcal{Q}_{\mathrm{P}}$ to the total flow
	rate when a negative pressure drop is imposed.
	
	To what extent does this cylinder approximation remain valid? The regime of
	validity can be ascertained by comparing with the numerical results. The calculations from Eq.~\eqref{dp2}
	[shown by open squares] are included in Fig.~\ref{fig:dp_vs_Q}. By its definition, at $\Delta p = 0$
	the cylinder approximation provides the exact (here negative) value of $\mathcal{Q} \equiv \mathcal{Q}_{\mathrm{DO}}$,
	but at different fluid flow rates there is some discrepancy, especially where $|\mathcal{Q} -\mathcal{Q}_{\mathrm{DO}}|$
	is higher. Nevertheless, these results demonstrate that the effective cylinder is quite a sensible
	approximation even when $\mathcal{Q}$ and $|\Delta p|$ become very large. For example, the leftmost and
	rightmost squares for the contracting channel give a relative deviation in $|\Delta p|$ of about $2$
	and $10\%$, respectively, and the other channels show much better agreement with the numerical curves.

	\section{Concluding remarks}\label{sec:conclusion}
	
	The results we have presented here complement those of our earlier study of
	planar slits~\cite{asmolov.es:2025,silkina.ef:2026} and extend many of the
	conclusions drawn there to axisymmetric channels, which are considered to be
	more realistic models for many systems, including artificial nanotubes,
	real porous membranes, and biological pores. Such a geometry generally hinders
	analytical progress, and we were driven to numerical solutions at an earlier stage
	than for slits. Nevertheless, the results suggested a powerful approximation
	that can easily be handled, making it suitable for both interpreting numerical
	and experimental data and for predictive purposes.
	
	We found that the diffusio-osmotic flow rate $\mathcal{Q}_{\mathrm{DO}}$ in cylinders
	is identical to that in planar slits. However, in channels with variable
	cross-sections, it is different. This gives them a special role, as the
	diffusioosmosis is set up also by their geometry, which adds a new dimension
	to the problem.
	
	Given that the shape contribution has a significant effect on the value of
	$\mathcal{Q}_{\mathrm{DO}}$, it is natural to enquire what happens to $\mathcal{Q}_{\mathrm{P}}$.
	We have investigated the role of the pressure drop as an additional driving
	force for the fluid flow rate $\mathcal{Q}$. It has been found that whilst
	at a non-zero pressure drop the flow in cylinders and slits has many features
	in common, there is a quantitative difference. As the pressure drop is
	decreased/increased, $\mathcal{Q}$ is shifted to larger/smaller values,
	but the shift for a cylinder is smaller than that appropriate to a planar slit.
	In the case of the cylinder, it was possible to set $\Delta p$ to obtain $\mathcal{Q}$.
	For channels with variable cross-sections, this is not straightforward,
	so we have set $\mathcal{Q}$ to find $\Delta p$. However, the curves
	$\Delta p (\mathcal{Q})$ remain quite similar to those obtained for cylinders.
	Namely, they are close to linear, although the slope now depends on the
	channel shape. The description of these curves is then closely akin to that of
	the cylinder treatment, provided the correct effective radii (related to the
	channel geometry) are chosen. This is confirmed by the results of the cylinder approximation discussed in
	Sec.~\ref{sec:discussion} and included in Fig.~\ref{fig:dp_vs_Q}. An important
	prediction of this simple approximate formula is also that it clarifies how
	the channel shape (together with a relevant $\Delta p$) acts to increase or decrease
	the fluid flow rate compared to that in a true cylinder. While the non-uniformity
	in the pore geometry in a real material should presumably smear out these effects,
	they can definitely be exploited in various lab-on-a-chip devices with uniform
	channels of known size and form.
	
	Several of our theoretical predictions, including the cylinder approximation,
	could be tested experimentally. The diffusio-osmotic flow rate $\mathcal{Q}_\mathrm{DO}$
	can be measured using the technique developed by \citet{lee.c:2014} for rectangular nanochannels.
	Although we are not aware of such measurements in axisymmetric channels with $\Delta p \neq 0$,
	these should, in principle, be feasible. Ideally, a theoretician would prefer to
	have the flow rate data for a material with well-defined pores of relevant dimensions
	and shape, rather than for poorly characterized ones. Nowadays, suitable materials
	can already be fabricated, although this remains a tall order.
	If the confining geometry is known, the hydrodynamic resistivity $\mathcal{W}$,
	which controls the slope of the $\Delta p$ vs $\mathcal{Q}$ curve, can be readily
	calculated analytically. Otherwise, one can infer it from an additional
	experiment on a pressure-driven flow with $\Delta c=0$. Once $\mathcal{W}$ is determined,
	$\mathcal{Q}$ can immediately be obtained as described in Sec.~\ref{sec:discussion}
	and compared with the measured one. It would be of some interest to examine
	the accuracy of the cylinder approximation in such an experiment.

	Finally, we have stressed the connection between the flow rate $\mathcal{Q}$
	and the ionic flux $\mathcal{J}$. This connection is usually swept under the
	carpet, but is highly relevant to understanding problems related to desalination.
	It is commonly considered that inducing a fluid flow toward the low-concentration
	side permits the extraction of fresh water from seawater by concentrating the latter.
	Our results suggest that the diffusio-osmotic flow in a thick axisymmetric channel
	always tends to equalize the salt concentrations on the two sides, and that
	this behavior cannot be affected by external hydrostatic pressure or channel
	geometry. This follows from the coupling between the fluid flow rate $\mathcal{Q}$
	and the ionic flux $\mathcal{J}$.
	Figures~\ref{fig:j_cyl} and \ref{fig:j_pe} clearly illustrate that when
	$\mathcal{Q}$ is negative, the ionic flux has the same sign. At first sight,
	it appears that diffusioosmosis is unlikely to be suitable for desalination.
	However, this conclusion refers to thick channels only and may change for
	thinner ones with overlapping diffuse layers. It would be of much interest
	to extend our approach to this case.

	It is likely that a detailed comparison of theory and experiment for these
	systems will soon be possible, providing a searching test of the ideas and
	results presented here. These can be used for the interpretation of
	diffusioosmosis in single channels and porous media, as well as guiding
	the design of micro- and nanofluidic devices.
	
	\begin{acknowledgments}
		
		This work was supported by the Ministry of Science and Higher Education of the Russian Federation.
	\end{acknowledgments}
	
	\appendix
	\section{Constraints of the Linear Theory: A Hint for Experimentalists}\label{a:1}
	
	As noted in Sec.~\ref{sec:global}, the classical Onsager matrix approach fails to capture the transport
	phenomena under realistic concentration drops. It is of considerable interest to evaluate quantitatively its applicability limits.

\begin{figure}[h]
		\centering
		\includegraphics[width=1.0\columnwidth]{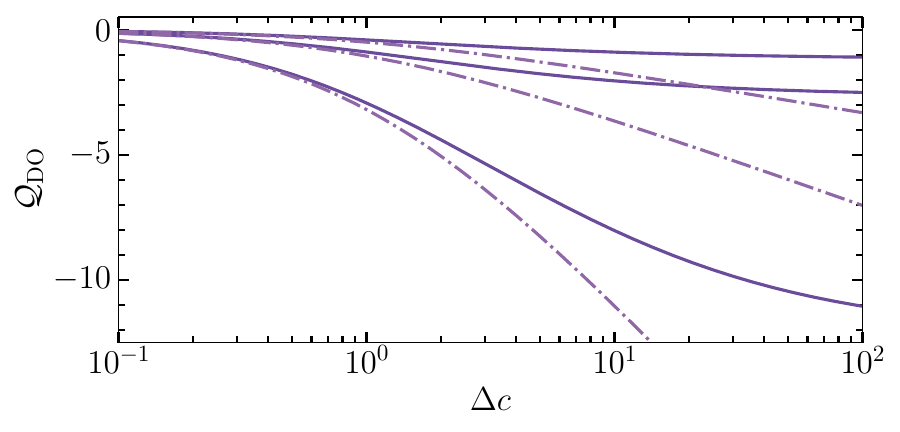}
		\caption{Diffusio-osmotic fluid flow rate $\mathcal{Q}_{\mathrm{DO}}$ in a cylinder as a function
			of the concentration drop $\Delta c$, calculated for several values of the
			Gouy-Chapman length: $\ell_{GC} = -1$, $-5$, and $-10$~nm (from bottom to top).
			Solid curves correspond to the exact numerical integration, while dash-dotted
			curves represent analytical predictions of Eq.~\eqref{eq:Q_DO_linear}.}
		\label{fig:appendix_graph}
	\end{figure}

	To illustrate the exact boundary where the linear approximation breaks down,
	in Fig.~\ref{fig:appendix_graph} we present the simplest example of $\mathcal{Q}_{\mathrm{DO}}$
	in a cylinder as a function of the concentration drop $\Delta c$ computed for several
	negative $\ell_{GC}$ (solid curves). Also included are the diffusio-osmotic flow rates
	calculated within a standard linear framework (dash-dotted curves), which implies
	\begin{equation}
		\mathcal{Q}_{\mathrm{DO}} \simeq - \ln (1+\Delta c) \left[ \beta \phi_{s}(0) + 4 \ln \left( \cosh \frac{\phi_{s}(0)}{4} \right) \right].
		\label{eq:Q_DO_linear}
	\end{equation}
	It can be seen that the computed results are irreconcilable with Eq.~\eqref{eq:Q_DO_linear}.
	This equation overestimates $|\mathcal{Q}_{\mathrm{DO}}|$, and provides a reasonable fit to the
	numerical data only up to $\Delta c \simeq 1$, which is equivalent to the two-fold
	concentration ratio in reservoirs.
	
	Application of the linear theory to fit experimental data at higher $\Delta c$
	inevitably yields a higher $|\phi_s|$ (deduced from the measured $\mathcal{Q}_{\mathrm{DO}}$)
	than its actual value. Conversely, if $\phi_s$ is known, the linear theory
	would predict a larger $|\mathcal{Q}_{\mathrm{DO}}|$ than is generated.

	This ``two-fold ratio'' constraint is particularly critical for practical energy applications,
	such as blue energy harvesting based on salinity gradients. In realistic systems
	involving river- and seawater, the concentration ratio routinely reaches several
	orders of magnitude (up to a factor of $10^3$ or higher). Given that the linear framework
	fails even for a modest concentration drop, applying linear Onsager relations
	to real-world membrane devices is fundamentally inadequate and leads to a severe
	underestimation of non-linear transport phenomena.
	
	\section*{DATA AVAILABILITY}
	
	The data that support the findings of this study are available within the
	article.
	
	\section*{AUTHOR DECLARATIONS}
	
	The authors have no conflicts to disclose.

\end{document}